\documentclass[12pt]{article}

\usepackage{natbib}

\usepackage{amsmath,amssymb}

\RequirePackage[colorlinks,citecolor=blue,urlcolor=blue]{hyperref}

\usepackage{times}
\usepackage{geometry}
\usepackage[nolists,figuresonly]{endfloat}

\renewcommand{\baselinestretch}{1.7} \rm

\usepackage{bm} 

\usepackage{graphicx}
\graphicspath{{./figures/}} 

\usepackage{tikz}
\usetikzlibrary{fit,positioning,calc}
\usetikzlibrary{plotmarks}

\newcommand{\fourall}{\text{ }\forall\text{ }}

\newcommand{\dotr}{\mbox{$\boldsymbol{\cdot}$}}

\begin{document}

\centerline{\LARGE\textbf{Gaussian Tree Constraints Applied to}}
\centerline{\LARGE\textbf{Acoustic Linguistic Functional Data}}
\centerline{\normalsize{Nathaniel \textsc{Shiers}, John A. D. \textsc{Aston}, Jim Q. \textsc{Smith}, and John S. \textsc{Coleman}}}

\noindent\rule{\textwidth}{0.4mm}

\normalsize{
\noindent Evolutionary models of languages are usually considered to take the form of trees. With the development of so-called tree constraints the plausibility of the tree model assumptions can be addressed by checking whether the moments of observed variables lie within regions consistent with trees. In our linguistic application, the data set comprises acoustic samples (audio recordings) from speakers of five Romance languages or dialects. We wish to assess these functional data for compatibility with a hereditary tree model at the language level. A novel combination of canonical function analysis (CFA) with a separable covariance structure provides a method for generating a representative basis for the data. This resulting basis is formed of components which emphasize language differences whilst maintaining the integrity of the observational language-groupings. A previously unexploited Gaussian tree constraint is then applied to component-by-component projections of the data to investigate adherence to an evolutionary tree. The results indicate that while a tree model is unlikely to be suitable for modeling all aspects of the acoustic linguistic data, certain features of the spoken Romance languages highlighted by the separable-CFA basis may indeed be suitably modeled as a tree.

\medskip

}

\noindent\rule{\textwidth}{0.4mm}


%
%
%

\newpage

\section{INTRODUCTION}
\label{sec:intro}

\let\thefootnote\relax\footnote{Nathaniel Shiers is PhD Candidate, Department of Statistics, University of Warwick, Coventry, UK (E-mail: n.l.shiers@warwick.ac.uk). John A. D. Aston is Professor, Statistical Laboratory, University of Cambridge, Cambridge, UK (E-mail: j.aston@statslab.cam.ac.uk). Jim Q. Smith is Professor, Department of Statistics, University of Warwick, Coventry, UK (E-mail: j.q.smith@warwick.ac.uk). John S. Coleman is Professor, Phonetics Laboratory, University of Oxford, UK (E-mail: john.coleman@phon.ox.ac.uk). The first author acknowledges the support of ESRC grant ES/I90427/1. The second author acknowledges the support of EPSRC grant EP/K021672/1. The authors wish to thank Piotr Zwiernik for very helpful discussions. They are also very much indebted to Pantelis Hadjipantelis for preprocessing the Romance language data set.}

As interest in functional data analysis (FDA) has increased, so has the development of theory \citep{ram05fda,hor12inffda} and the range of applications (e.g.\ brain imaging (\citet{sor13ima}), climatology (\citet{bes00aff}), and medical research (\cite{rat02fda})). This has been facilitated in part through  better access to functional data and through greater availability of computational power for analyses. The use of FDA in statistical phonetics has recently attracted attention (e.g.\ \citet{koe08spv, moo10alm}). Such analyses, which involve acoustic functional data, have provided particularly promising and interesting results in a diverse range of settings. For example, \cite{gra07cil} uses a polynomial basis expansion to examine pitch variation in English, while \citet{ast10lpa} investigates Qiang, a Sino-Tibetan language, and finds previously unidentified gender differences amongst speakers via a functional principal components based modeling approach.

The acoustic structure of spoken words can be used to investigate areas of linguistic interest in a similar way that discrete (alphabetic) representations of speech have been utilized to make cross-language comparisons, and more recently inferences regarding proto languages (\citet{hoc91phl},\citet{bou13ara}
). The differences and similarities between spoken languages suggest that any meaningful functional observations taken across languages are unlikely to be independently, identically distributed. As such it is probable that the language relationships form a tree or network structure, which may be informative about possible historical developments of these languages. If this alternative (acoustic) approach can be used to corroborate known and uncontroversial language relationships, then our methods offer great potential for less certain language relationships. For instance, this would be useful for languages where there are few historical records but in which inference of a family tree is reasonably supported by the contemporary data (e.g. African language families), or alternatively, in cases where reconstruction of a family tree is disputed, such as Greenberg's classification of native American languages (\citet{bol04pug}).

Relationships between languages have long been described as phylogenetic trees constructed using linguistic factors (e.g.\ \citet{sch60dds}) where all non-leaf variables are unobserved and represent features of the past languages before their divergence. \cite{qam} developed some of the first quantitative methods which were used to investigate evolutionary relationships between languages. More recently there have been large scale attempts to reconstruct trees or networks of languages (e.g.\ \citet{ppn} for the Indo-European language family, \citet{pms} for the Semitic language family). Some researchers have shifted away from describing the evolutionary language relationships via trees toward using networks (for example, \citet{tpc}, \citet{nuh}). 
However, trees have a somewhat more natural interpretation in terms of evolutionary structure, and assessing the suitability of a tree model for language data would therefore be of interest to researchers in linguistics. The focus of this paper is to examine functional acoustic data from speakers of five Romance languages (French, Italian, Portuguese, American Spanish, and Iberian Spanish) to provide insight at an exploratory level as to whether a tree may be adequate for describing certain features of these language relationships.

To address questions of tree-amenability we appeal to the notion of tree constraints. The theory of tree constraints is embedded in the area of algebraic statistics, a field which has a significant recent literature related to phylogenetics (e.g.\ \citet{stu05tip}, \citet{all08piv}). It has been known for some time that covariance functions of data on observed 
variables respecting an evolutionary tree 
must obey particular algebraic and semi-algebraic constraints, e.g.\ \citet{set00gmc}. Recently these have become much better understood (for example \citet{drt07asm}, \citet{all09ipl}, \citet{all12sdg}) and fully characterized in some cases (e.g.\ the binary case \citet{zwi11iic}, \citet{zwi12tcg}). In this paper, building on developments in \citet{shi12gid}, a Gaussian analogue of the binary tree constraints is applied to the covariances of component-by-component projections of the data. By considering the data component-wise, a more realistic and nuanced analysis can be performed which permits some observed features of linguistic data to be tree-amenable and indicates that that for others that any evolutionary tree is unlikely to provide a good explanation.

Section~\ref{sec:data} describes the data set and the preprocessing in preparing an audio recording for FDA. In a similar spirit to the work on object data by \cite{wan07ood}, for the application to Romance languages presented here, the observation units of interest are two-dimensional functional data objects known as spectrograms (time-frequency descriptions of the data). These are formed from transformations of audio recordings of people speaking single words. When regarded as functional data, observations are in fact stored as high dimensional objects. Therefore, in Section~\ref{sec:proj} tools from FDA are employed to transform high dimensional speaker data to a lower dimension. This is achieved through the novel approach of using between-language covariance as described in canonical function analysis (CFA) and combining it with a tensor decomposable covariance structure.

Having achieved the required reduction in dimension, Section~\ref{sec:tree} provides a brief summary of tree constraints. A fundamental but yet to be exploited constraint for use with Gaussian data is then introduced. Statistics associated with the violation of or adherence to the constraint are then constructed from the acoustic language data to answer the question of tree suitability. In Section~\ref{sec:treeapp}, in preparation for use with this Gaussian tree constraint, we describe the construction of a between-language cross-covariance matrix using the scores (projections) of the acoustic data. In Section~\ref{sec:sim}, the general effectiveness of the Gaussian tree constraint is investigated via simulations to assess its ability to correctly accept or reject tree-amenability. Section~\ref{sec:furthersim} entails further simulation, tailored to mimic aspects of the acoustic data, so the results are more immediately relevant to the setting. Section~\ref{sec:apply} addresses the ultimate aim of the study: to determine whether any evolutionary relationships between the Romance words represented in the acoustic data set can be described by a Gaussian evolutionary tree model. These tree constraints are then used to explore tree-amenability for each of the components of the chosen basis. It is found that a subset of the components adhere to the tree constraint. This suggests that some features of the acoustic linguistic data which distinguish between languages could have evolved in a tree-like manner whilst others have not. This is consistent with the current understanding that the development of Romance languages has been complex, involving much cross-language interaction (\citet{har88trl}) in addition to a historically well-documented common origin from Latin. Thus, attempting to fit a tree model to the entire data set is likely to be misguided based on the empirical evidence presented here. More appropriately, a different model can be fitted for each component, using the tree constraints to indicate whether to restrict the space of models to trees.

\section{ACOUSTIC FUNCTIONAL DATA SET}
\label{sec:data}

The data set of interest comprises audio recordings originating from speakers of one of five different Romance languages: French, Italian, Portuguese, Spanish (American), and Spanish (Iberian) - while two dialects of Spanish are being used in this study, they are treated as different spoken languages in this analysis as we are interested in pronunciation rather than textual representation, the difference between ``dialect'' and ``language'' being a matter of degree of difference rather than an absolute quantitative difference. Each recording is of some individual saying an integer from `one' to `ten' in their particular language. Recordings were made at a rate of 16000 samples per second and a resolution of 16 bits. In total there are 219 word recordings and each can be classified by the language, the gender of the speaker and the number being spoken. 
Observations of the same word being spoken in different languages are treated as sharing the same word attribute. For example the word `four' includes recordings of `quatre' (French) and `quattro' (Italian) as well the word `four' in other languages. 
Integers were chosen because these have no ambiguity in terms of translation making comparison of their use across languages straightforward. Furthermore, the cardinals `one' to `ten' of Romance languages (among many other words) stem from shared Latin forms (\citet{gla92ien}). This suggests that these words might also be suitable when comparing languages acoustically.

In this paper the observations are modeled as functional data as is becoming increasingly common in studies involving sound recordings (e.g.\ \citet{hol10mcp}). Such models make the reasonable assumption that the data have been obtained by observing an underlying function at finitely many discrete points along a continuum, and that this underlying function is smooth (i.e.\ a certain number of derivatives exist).

\begin{figure}[htp]
\centering
\includegraphics[width=0.6\textwidth, height=0.6\textheight,keepaspectratio]{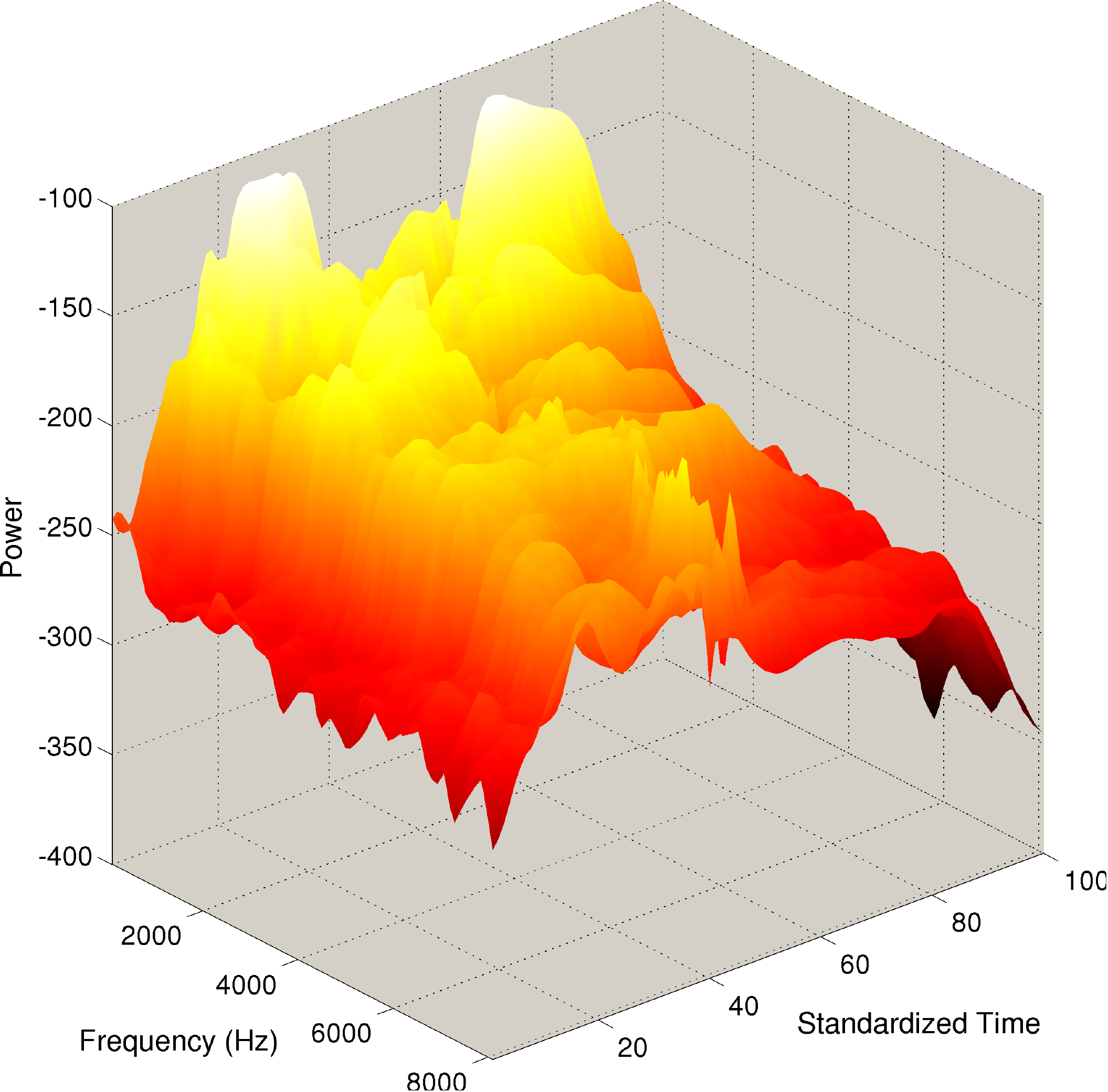}
\caption{Post-registration spectrogram of female French speaker saying `quatre'. It can be seen that there is greater power in the lower frequencies, and that the very beginning and end of the word are unsurprisingly two of the quietest regions.}
\label{fig:postwarpeg}
\end{figure}

The overall duration of a word can vary significantly per speaker as can the timings of intra-word elements (for instance syllables). Thus, to adjust for these differences all observations within a word grouping undergo registration (also known as alignment or warping, see \citet{ram05fda, luc00tnv}). A short-time Fourier transform is taken of each audio recording to produce a spectrogram. A spectrogram is a two-dimensional representation of audio signal energy intensity in frequency-time space (\citet{ful11ssa}). Spectrograms are a natural choice for representing power with functional data, though approaches such as Mel-frequency cepstra (\citet{dav80cpr}) can provide possible alternative representations. As part of the preprocessing, the standardization of word duration results in the time dimension being measured in generic time units. The value stored at a frequency-time point is a function of the power (or amplitude), with frequencies binned every 100Hz up to the Nyquist frequency of 8000Hz. The resulting spectrograms are stored as matrices of 81 frequency by 100 time points. Figure~\ref{fig:postwarpeg} is the spectrogram of a female French speaker saying the word `quatre'. Broadly, this interpolated plot indicates that there is greater power in the lower frequencies, and that the beginning and the end portions of the standardized time period are quieter.

\subsection{Notation}

The underlying function of each spectrogram is denoted $x_{l,m}^{d,g}(f,t)$ with the two dimensions $f$ and $t$ referring to frequency and time respectively. Recall that each spectrogram is derived from a spoken word - the subscripts and superscripts encode observational information: $l = 1, \ldots, n_l$ denotes the language being spoken; $d = 1, \ldots, n_d$ indicates the word being spoken; $m = 1, \ldots, m_{ld}$ is a counter where $m_{ld}$ is the number of observations of word $d$ from language $l$; $g$ refers to the gender of the speaker. 

It is well documented that there are differences in the acoustics of male and female speakers which go beyond a simple shift in the spoken frequencies. For instance \citet{nit90amm,pep13vsg}. 
\citet{par96lig} present a statistical method for discriminating between speaker gender of short acoustic recordings. In their analysis of seven Indo-European languages (of which Romance is a subset), gender was correctly identified on average 98\% of the time. This suggests that there are commonalities in acoustic gender differences across Indo-European languages. In light of this result, it is judged that gender should be adjusted for at the macro level: $$x_{l,m}^{d}(f,t) = x_{l,m}^{d,g}(f,t) + \tilde{x}^{g}(f,t)$$ where $\tilde{x}^{g}$ is the difference between the mean of all samples with gender $g$ and the mean of all samples. Henceforth it will be the gender adjusted function that will be the object of interest in this paper.

The mean spectrograms for language $l$, word $d$ are defined in (\ref{eq:gpfuncmeanij}), for language $l$ in (\ref{eq:gpfuncmeani}), and the grand mean spectrogram in (\ref{eq:gpfuncmean}).
\begin{align}
\bar{x}_{l}^{d}(f,t)&=\frac{1}{m_{ld}}\sum_{m=1}^{m_{ld}}x^{d}_{l,m}(f,t) \label{eq:gpfuncmeanij}\\
\bar{x}_{l}(f,t)&=\frac{1}{m_{l\dotr}}\sum_{d=1}^{n_d}m_{ld}\bar{x}_{l}^{d}(f,t) \label{eq:gpfuncmeani}\\
\bar{x}(f,t)&=\frac{1}{m_{\dotr\dotr}}\sum_{l=1}^{n_l}m_{l\dotr}\bar{x}_{l}(f,t) \label{eq:gpfuncmean}
\end{align}
where $m_{l\dotr} = \sum_{d=1}^{n_d}m_{ld}$, $m_{\dotr\dotr} = n = \sum_{l=1}^{n_l}m_{l\dotr}$, and for $t \in \mathcal{T}$, $f \in \mathcal{F}$. The parameters $m_{\dotr\dotr}$ and $n$ will be used interchangeably depending on whether summation is being emphasized.

\section{GROUP BASED PROJECTIONS OF FUNCTIONAL DATA}
\label{sec:proj}

Dimension reduction is a well-studied area of statistics with tools such as principal component analysis (PCA) and  multidimensional scaling (e.g.\ see \citet{jol02pca,cox10mds}) having widespread use. Functional counterparts of such techniques have also been formulated, for example functional principal component analysis (FPCA) (\citet{Castro86,Rice91,yao06psm}), and functional multidimensional scaling (\citet{miz06grf}). 

The acoustic data presented in this paper benefit from a dimension reduction in two main ways. First and foremost, dimension reduction provides a route to feature extraction whilst also reducing unwanted noise. Second, if subsequent to the reduction it is found that $n \geq p$ then techniques which make use of inverse covariances can be implemented straightforwardly. If this is not so, standard estimates will produce singular sample covariance matrices. 
Of course these benefits must be balanced against potential information loss from the data reduction. One approach to feature extraction which mitigates against this loss is to find an ordered basis which prioritizes one or more characteristics of interest. Thus by projecting data onto the first few components of such a basis the most prominent aspects of the data are retained whilst what remains is treated as noise. In the cases of PCA and FPCA, the dimension reduction is optimized so as to efficiently capture modes of variation. Such techniques are often used in linguistic and semantic analyses, for example \citet{lee01rne} and \citet{wen10stm}.
However, as our focus is on macro-language comparisons, we argue that the feature of interest is the between- to within-language covariance, and it is this which should directly inform the method selected to construct a basis. When data is known a priori to be grouped then canonical function analysis (CFA) and its multivariate analogue canonical variate analysis (CVA) are standard techniques implemented to select variables to discriminate between groups. These tools are therefore the starting points for our analysis.

\subsection{Canonical Function Analysis}
\label{sec:cfa}

Here we present CFA as a tool for FDA to produce a basis which maximizes between- to within-group variation (subject to the basis component functions being uncorrelated) with the intention of achieving an efficient dimension reduction. The finer details of CFA can be found in \citet{kii92cva}. To illustrate CFA, consider a set of one-dimensional functional data 
denoting the between-group covariance function as $B(u_{1},u_{2})$ and within-group covariance function as $W(u_{1},u_{2})$ (with $u_{1}, u_{2} \in \mathcal{U}$, $\mathcal{U} \subset \mathbb{R}$) 
where both functions are considered to be bounded and piecewise continuous.
The aim is to identify canonical functions $h_{q}(u)$ such that between-group variation is maximized relative to within-group variation under the restriction that each canonical function is uncorrelated to every other. This is expressed as a generalized eigen-equation which simplifies to:
\begin{equation}
\intop_{\mathcal{U}}(B(u_{1},u_{2})-\,\lambda_{q}W(u_{1},u_{2}))h_{q}(u_{2})\,du_{1}=0\label{eq:optfunc}
\end{equation}
with eigenvalues $\lambda_{q}$ and eigenfunctions $h_{q}$.
There may be countably infinite solutions to equation (\ref{eq:optfunc}) but in discretized estimation, only a maximum of s (say) will have non-zero $\lambda_{q}$. Pairs of canonical functions and real numbers $(h_{1}(u),\lambda_{1}),\ldots,(h_{s}(u),\lambda_{s})$ can be found by solving $(\ref{eq:optfunc})$ numerically, where $\lambda_{1}, \ldots, \lambda_{s}$ is a monotone decreasing sequence. 
An $r$-dimensional projection of the data is obtained using the first $r$ canonical functions, and this projection is such that the between- to within-group covariance is maximally retained. 

\subsection{Separable Covariance Functions}
\label{sec:covfuncsep}
Recall that the Romance data set comprises spectrograms which have time and frequency directions. A straightforward model for describing how these directions interact is that of separable covariance. The assumption underpinning this model can be encapsulated as there being no dependency between the (standardized) time and frequency of the data. This is, of course, a significant simplification of the likely underlying model. However, particularly for computational considerations, this assumption can very useful.
Recall that a covariance function $C$ is said to be separable if:
$$C((f_1,t_1),(f_2,t_2)) = C_f(f_1,f_2)C_t(t_1,t_2)$$
where $C_f$ and $C_t$ are functions only of their arguments. The factored covariances provide an understanding of how frequency or time dimensions of the spectrograms vary when the other has been averaged out. 
However, the main purpose of the assumption becomes apparent for the Romance data set subsequently (as described in Section~\ref{sec:covsep}) when use of the separable model of covariance overcomes the challenge of covariance rank deficiency.

Under the separable covariance assumption for two-dimensional data the CFA optimality equation equivalent to (\ref{eq:optfunc}) is:
\begin{equation}
\int\limits_{\mathcal{F}} \int\limits_{\mathcal{T}} (B_{f}(f_{1},f_{2}) B_{t}(t_{1},t_{2})-\lambda_{q}W_{f}(f_{1},f_{2}) W_{t}(t_{1},t_{2}))h_{q}(f_{2},t_{2}) \,dt_{2}\,df_{2}=0.
\label{eq:withassump}
\end{equation}
It can be easily shown that the solutions to this equation can be obtained as the product of the solutions to two CFAs performed on the frequency and time covariances separately. Thus given any canonical function pairs ($h_{q_{f}}(f_{2}),\lambda_{q_{f}}$) and ($h_{q_{t}}(t_{2}),\lambda_{q_{t}}$) from a frequency and time CFA respectively, the products provide a solution to (\ref{eq:withassump}): $h_{q}(f_{2},t_{2}) = h_{q_{f}}(f_{2})h_{q_{t}}(t_{2})$ and $\lambda_{q} = \lambda_{q_{f}}\lambda_{q_{t}}$. Moreover, any solution to (\ref{eq:withassump}) can be obtained from such products. This result is useful when proceeding to obtain numerical solutions to (\ref{eq:withassump}).

\subsection{Canonical Variate Analysis}
\label{sec:cva}

Although less frequently implemented than PCA, the theory of CVA as a multivariate tool has been well developed in \citet[Chapter 11]{krz90pma}. However, beyond being a purely multivariate technique, CVA can also be used with functional data as an approximation to CFA as is presented in \citet{kii92cva}. The technicalities of implementing CVA do not differ whether in a functional or multivariate setting, although it is sometimes necessary to interpret their outputs differently, as is encountered with other tools (e.g. \citet{fer12ima}).

In practice spectrograms are often discretized representations of underlying functions, and so each function $x_{l,m}^{d}$ is instead given by a matrix $\bm{X}_{l,m}^{d}$ with time-frequency dimensions $n_f \times n_t$ (i.e. the number of sample points of the frequency and time). These finite approximations tend to be high dimensional and so the question of dimension reduction is pertinent. By concatenating the rows of these matrix representations of spectrograms the data corresponds to the vector description of CVA. 
As CVA considers each covariance entry independently of its adjacent values, this does not affect the implementation of CVA. The only notable downside of concatenation is that it can obscure visual representation and description of the data.

Recall that the aim of CVA is to find successive uncorrelated vectors $\bm{a}$ that form linear combinations $y=\bm{a}\bm{x}^{T}$ (where $\bm{x}$ is $p$-dimensional data) that maximize the ratio 
of the between-groups covariance ($\bm{B}$) to the within-groups covariance ($\bm{W}$). 
In the context of the paper, $\bm{B}$ describes the variation between the per-language mean spectrograms and the grand mean spectrogram, whereas the $\bm{W}$ describes the variation between individual observations and the associated per-language mean spectrograms.

Finding the optimal $\bm{a}$ is equivalent to solving $(\bm{W}^{-1}\bm{B}-\lambda\bm{I})\bm{a}^{T}=\bm{0}$ where $\lambda\in \mathbb{R}$. This reduces to performing an eigenanalysis on $\bm{W}^{-1}\bm{B}$ - the eigenvector corresponding to the largest eigenvalue is the optimal $\bm{a}$. 
As with CFA, canonical pairs $(\bm{a}_{r}^{T},\lambda_{r})$ are sought. These are found through a full eigenanalysis of $\bm{W}^{-1}\bm{B}$ such that $\lambda_{1}>\lambda_{2}>\ldots>\lambda_{s}>0$ where $s=\min(p,n_l-1)$ is the number of non-zero eigenvalues of $\bm{W}^{-1}\bm{B}$. Thus $(\bm{a}_{r}^{T},\lambda_{r})$ produces the $r$th greatest ratio of between- to within-language variability. Hence, the optimal projection to $r$ dimensions requires only $(\bm{a}_{1}^{T},\lambda_{1}),\ldots,(\bm{a}_{r}^{T},\lambda_{r})$. Theoretically, CVA is designed for use with Gaussian data and assumes that within-group covariance is equal across groups. If these assumptions hold true then CVA is optimal for identifying modes of variability which distinguish between groups.

\subsubsection{Separable-CVA}
\label{sec:covsep}

As mentioned in Section~\ref{sec:covfuncsep}, the overall solutions to a CFA optimality problem with a separable covariance structure can be found as the product of solutions to CFAs of the decomposed covariance functions. We propose combining a tensor decomposable covariance structure with CVA in order to obtain numerical solutions to the decomposition of the separable-CFAs. This, when taking products, also gives solutions to the overall CFA. While separable covariance structures have been adopted elsewhere in the literature (e.g.\ \citet{jon11eif, ast12esc}), this is a novel approach for both CVA and CFA. Even though the assumption behind separable covariance is strong, the accuracy of the assumption for CVA and CFA only impacts on basis efficiency not basis validity. If the data is far from separable, then simply a higher number of dimensions will be needed to retain the same amount of information.

The main purpose of assuming a tensor-decomposable covariance structure is to overcome the obstacle of rank-deficient sample covariance matrices caused by the length of the observations exceeding the number of observations (i.e.\ $p > n$). This is not just a problem with the Romance speaker data set but is commonly encountered with functional data sets due to their often high-dimensionality (e.g.\ \citet{lon05nne}). Rank deficiency obstructs using CVA to obtain numerical solutions to CFA. Theoretically in CFA an inverse function $W^{-1}$ is neither required nor is usually bounded, whereas in CVA $\bm{W}^{-1}$ is needed for the eigenanalysis of $\bm{W}^{-1}\bm{B}$ but cannot be obtained because in this case $\bm{W}$ is singular.

In the observational matrix setting, $\bm{C}$ is separable if:
$$\bm{C}((f_1,t_1),(f_2,t_2)) = \bm{C}_f(f_1,f_2)\otimes\bm{C}_t(t_2,t_2)$$
where $\otimes$ is the standard Kronecker product. Using known results of the Kronecker product (see \citet{lan85tm-} for example), the separability assumption in the multivariate setting implies:
\begin{equation}
\bm{W}^{-1}\bm{B} = (\bm{W}_t^{-1}\otimes \bm{W}_f^{-1})(\bm{B}_t  \otimes \bm{B}_f) = \bm{W}_t^{-1}\bm{B}_t \otimes \bm{W}_f^{-1}\bm{B}_f
\label{eq:kron}
\end{equation}
where the estimates of separate within- and between-language covariance matrices in the frequency direction are:
\begin{align*}
\bm{\hat{B}}_f[f_1,f_2] = \frac{1}{n_l-1}&\sum_{l=1}^{n_l}\frac{m_{l\dotr}}{n_t}\sum_{t=1}^{n_t}\tilde{\bm{X}}_l[f_1,t]\tilde{\bm{X}}_l[f_2,t]\\
\bm{\hat{W}}_f[f_1,f_2] = \frac{1}{n-n_l}&\sum_{l=1}^{n_l}\sum_{d=1}^{n_d}\sum_{m=1}^{m_{ld}}\frac{1}{n_t}\sum_{t=1}^{n_t}\tilde{\bm{X}}_{l,m}^{d}[f_1,t]\tilde{\bm{X}}_{l,m}^{d}[f_2,t]
\end{align*}
where $\tilde{\bm{X}}_l[i,j] =\bar{\bm{X}}_l[i,j]-\bar{\bm{X}}[i,j]$ and $\tilde{\bm{X}}_{l,m}^{d}[i,j] = \bm{X}_{l,m}^{d}[i,j]-\bar{\bm{X}}_l[i,j]$ with equivalent estimates for the time direction. Treating each frequency and time sample as a separate observation leads to the product covariance matrices $\bm{W}$ and $\bm{B}$ having higher ranks than previously. Explicitly, for $\bm{W}^{-1} = (\bm{W}_f\otimes~\bm{W}_t)^{-1}$ to be nonsingular, we need that $nn_f \geq n_t$ and $nn_t \geq n_f$. This is equivalent to requiring $n \geq \tfrac{\max(n_f,n_t)}{\min(n_f,n_t)}$. This contrasts to the previous condition $n\geq p = n_fn_t$. So the new requirement is usually significantly more relaxed, and CVA can often then be implemented.

An eigenanalysis of $\bm{W}_{f}^{-1}\bm{B}_{f}$ produces eigenvalues $(\lambda_{f1}, \lambda_{f2}, \ldots, \lambda_{fn_f})$ and corresponding eigenvectors $(\bm{c}_{f1}, \ldots, \bm{c}_{fn_f})$ with equivalent output for the time covariances $\bm{W}_{t}^{-1}\bm{B}_{t}$. Sorting decreasingly, the vector obtained from $(\lambda_{f1}, \ldots, \lambda_{fn_f}) \otimes (\lambda_{t1}, \ldots, \lambda_{tn_t})$ is denoted $(\lambda_{1}, \lambda_{2}, \ldots, \lambda_{n_f n_t})$ and the Kronecker product of the corresponding eigenvectors results in matrices denoted $(\bm{c}_{1}, \bm{c}_{2}, \ldots, \bm{c}_{n_f n_t})$ of size $n_f \times n_t$, solving the overall CVA. It should be noted that while this basis defined is based on an assumption of separability, it nevertheless provides a complete basis of the space. So although when separability does not hold the basis is less efficient and is rather longer than it needs to be, the basis is still valid. For further details, see an analogous argument for separable PCA in \citet{ast12esc}.

\subsection{Application of CVA as an Approximation to CFA}
\label{sec:select}
As motivated, the separable-CVA is used to approximate the separable-CFA of the Romance languages data to achieve a dimension reduction based on components which maximize between- to within-language variability. This is a suitable approximation to make as the functional spectrograms have been sufficiently densely sampled during discretization.

\begin{figure}[htb]
\centering
\includegraphics[width=0.85\textwidth, height=0.85\textheight,keepaspectratio]{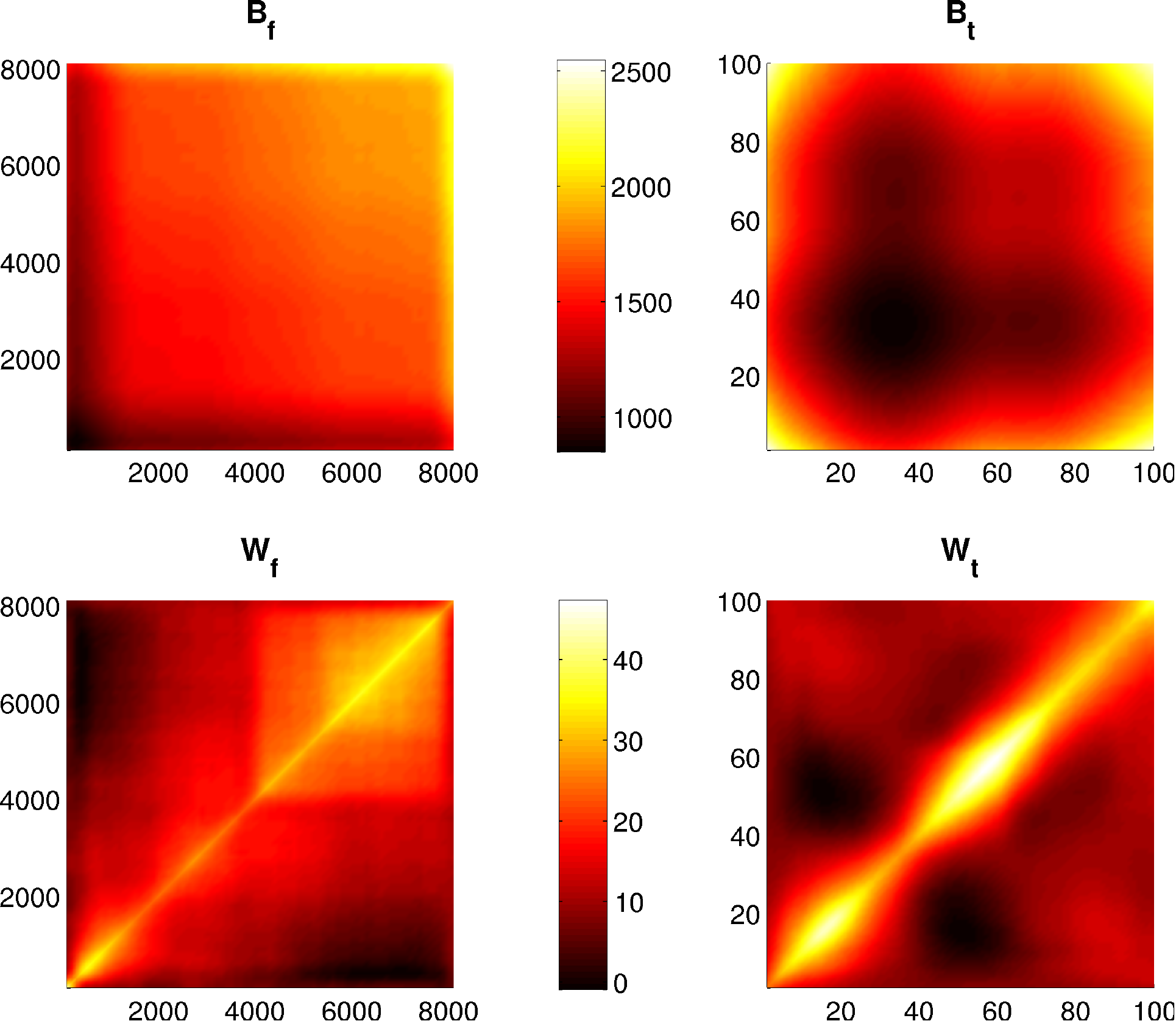}
\caption{Sample between-language and within-language covariances of speech data for frequency and time directions. There is a clear ridge along the diagonal of the within-group covariances indicating that similar times and frequencies are highly positively correlated. The higher correlations in the high frequencies of the within-group frequency covariance are associated with recordings at lower audio sampling rates capturing fewer details in these ranges of frequencies. Rerunning the analyses performed in Section~\ref{sec:apply} whilst excluding these higher frequencies produces broadly similar results.}
\label{fig:bwcovscrop}
\end{figure}

Interpolated plots of the between- and within-language covariances for both the time and frequency directions of the spectrograms are given in Figure~\ref{fig:bwcovscrop}. The $\bm{B}_f$ plot displays positive covariance which increases as the frequency increases. The time covariance $\bm{B}_t$ 
is positive throughout with an internal minimum and the highest covariances  corresponding to beginning and end time points. These particularly high time corner covariances could indicate genuine similarities at the beginning and ends of spoken words or could be an artefact of the spectrogram registration. 
The covariances $\bm{W}_t$ and $\bm{W}_f$ both have a ridge along the diagonal which reassuringly suggests that similar time and frequency points have strong covariances within languages.

\begin{figure}[htb]
\centering
\includegraphics[width=0.8\textwidth]{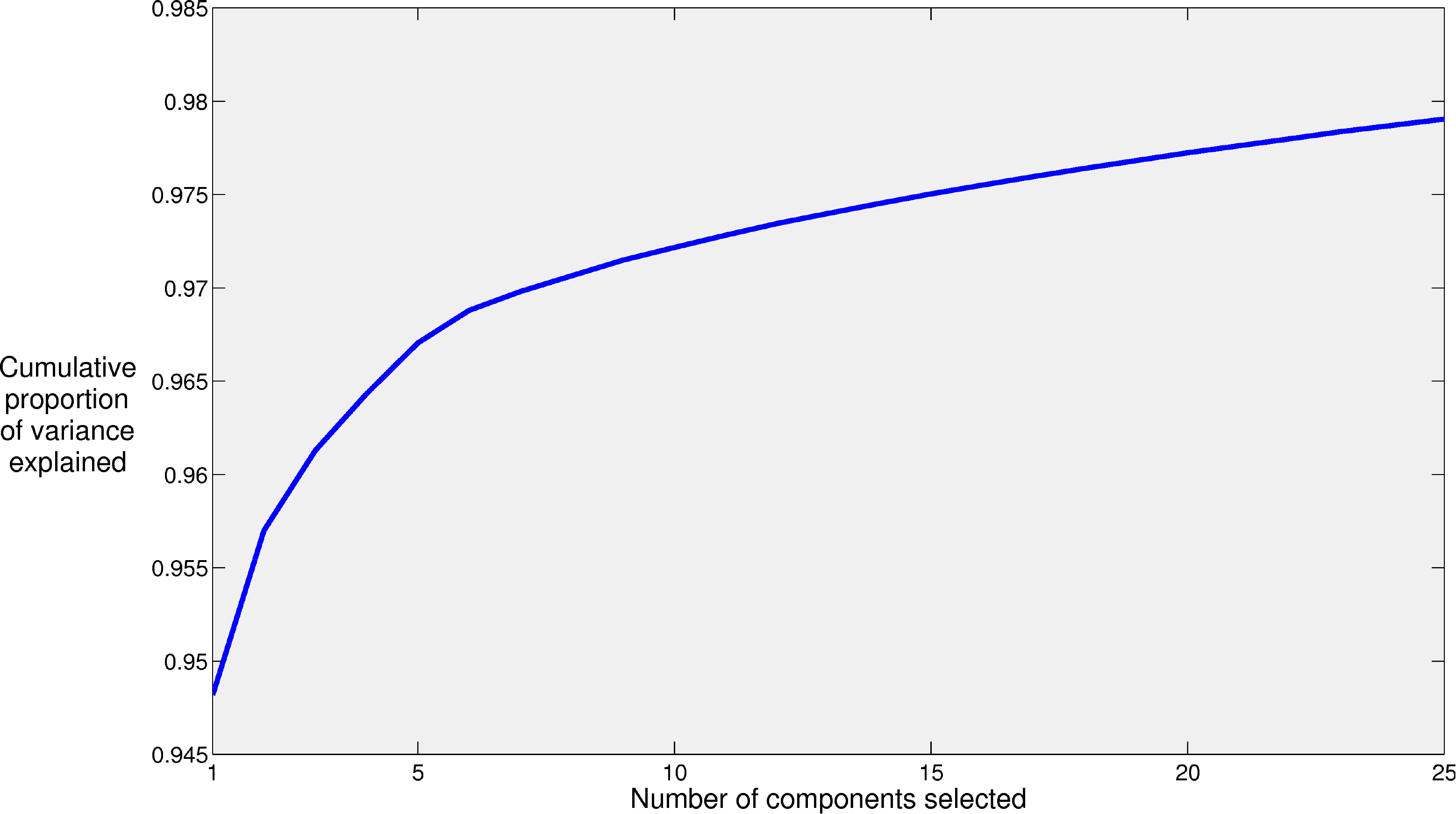}
\caption{Cumulative variation explained by number of components. The explanatory power of the first component in terms of between- to within-language variability is over 94\%.}
\label{fig:cumeig}
\end{figure}

When selecting a dimension $r$ to project to, it is unusual to have anything but an arbitrary albeit sensible method for selecting $r$. However, in some acoustic contexts (e.g.\ \citet{had12cff}) thresholds can be proposed based on sounds which are audible to humans. Otherwise, equivalent techniques to those employed with PCA (\citet{jol02pca}) can be used. 
For this linguistic study Figure~\ref{fig:cumeig} shows the cumulative variation explained by selecting particular numbers of components. This indicates that almost 95\% of the between- to within-language variance can be explained by a single component, and an additional two components take this figure to over 96\%.

Once a dimension $r$ has been selected, each observation from the language data can be projected into $r$-dimensional space: $\bm{Y} = \bm{A\tilde{\bm{X}}}^{T}$ where $\bm{A}$ is $r \times p$ with columns $\bm{c}_{1}, \ldots, \bm{c}_{r}$ and where $\bm{\tilde{X}}$ is $1 \times p$ and formed by concatenating the $n_f$ rows of length $n_t$ of the observation $\bm{X}$. The sub and superscripts of the observations are omitted solely for notational clarity.

\begin{figure}[!htb]
\centering
\includegraphics[width=1\textwidth, height=1\textheight,keepaspectratio]{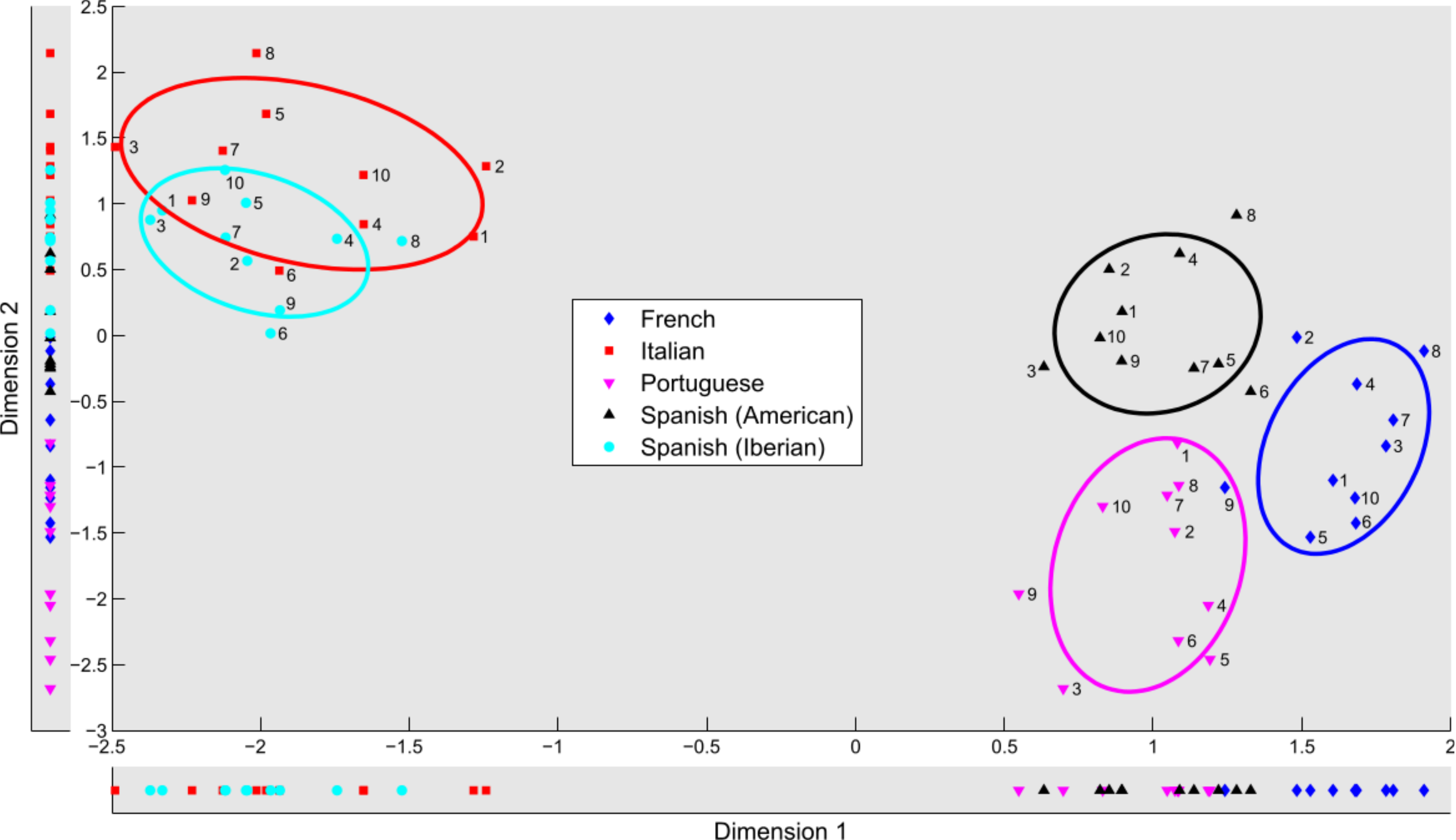}
\caption{Two dimensional separable-CVA projection of means of word observations with ellipses representing the standard deviation around the mean of plotted points. The second dimension is effective at distinguishing at the language level. The first dimension then discriminates between the Spanish dialects. As each component can relate to a different set of speech qualities, the proximities of the languages can differ depending on the dimension projection. For example, in the second dimension, the nasality of the speech appears to contribute to the separation of Italian and Portuguese. This is further explored in Section~\ref{sec:apply}.}
\label{fig:2dproj}
\end{figure}
To demonstrate the effectiveness of projection of the spectrograms in even two dimensions, the projections of the means of the word observations are plotted in Figure~\ref{fig:2dproj}. The results of the separable-CVA are encouraging as there are clear groupings of the projected word means from the same languages. Furthermore, the first two dimensions appear to reflect aspects known about the languages. The second dimension seems to distinguish the four distinct languages, and then the first dimension is able to separate the Spanish dialects (while also providing clear distinctions between some of the other languages). Given that the acoustic data are undoubtedly noisy, this indicates the effectiveness of separable-CVA at selecting components which discriminate on a group basis. Note that whilst CVA operates on all languages simultaneously rather than in a pairwise manner, this does not necessarily imply languages in close proximity post-projection share particular acoustic features. However, this can be examined in more detail, for example through studying the Hadamand matrices of each projection as is illustrated in Figure~\ref{fig:ftcomp2}.

\section[CONSTRAINTS AS DIAGNOSTICS FOR TREE-AMENABILITY]{CONSTRAINTS AS DIAGNOSTICS FOR\\ TREE-AMENABILITY}
\label{sec:tree}

The aim of this study is to examine the suitability of a tree for acoustic functional data. It is of particular interest to know whether there are certain features of these spoken Romance languages which could have developed over time in the manner of an evolutionary tree. CFA (or CVA) is compatible with this aim as it effectively identifies components with features which distinguish between languages. Thus a projection to $r$ dimensions provides $r$ different combinations of characteristics which could potentially be modeled as a tree. In order to help assess the plausibility of these $r$ hypotheses we turn to the topic of tree constraints. These are algebraic and semi-algebraic constraints which mathematically must be obeyed if a data set corresponds to the leaves of a Bayesian network which is a tree. Note here that the non-leaf vertices correspond to various unobserved ancestral languages. While there has been considerable recent interest in understanding graphical structures, including trees, in very general settings \citep{loh13sed}, these approaches are design for settings when all nodes (both internal and leaves) are observed. Here we are concerned with determining constraints when the internal nodes are unobserved. Examining whether these constraints are respected for a given data set determines whether that data set adheres to a tree structure (i.e.\ whether the data set is tree-amenable). Provided with $r$ component-by-component projections, a diagnostic for tree-amenability can be applied to each (see e.g.\ \citet{shi12gid} for a binary example). This can then be used as an exploratory tool to give insight as to whether the distinguishing characteristics captured by a component could have an evolutionary tree structure. We begin by giving a brief overview of the concept of tree constraints and how they form a natural choice for use with data sets such as the acoustic recordings in this study.

\subsection{Constraints on Observed Covariances Respecting Trees with Hidden Variables}

Tree constraints are useful for studies investigating evolutionary relationships between both observable and unobservable variables. These evolutionary relationships can be expressed graphically as phylogenetic trees where traditionally the objects of interest have been biological species (e.g.\ \citet{sip}), however, this idea naturally extends to other fields such as linguistics (e.g.\ \citet{dun05spr}). Phylogenetic trees are of particular statistical interest when formally considered as Bayesian networks describing conditional independence relationships (e.g.\ \citet{ifm}). In the case of discrete graphs in which all variables are assumed to be observed, \citet{loh13sed} make use of precision matrices to assess graph structure. However, typically Bayesian networks have both observed and hidden variables and these are much more complex to analyze. Furthermore, the data set in question is functional and thus the analyses are embedded in Gaussian assumptions and observations treated as Gaussian processes as opposed to individual points. This is where Gaussian tree constraints can provide useful insight into the difficult question of tree-amenability.

\begin{figure}[ht!]
\centering
\tikzstyle{vertex}=[circle,fill=black,minimum size=5pt,inner sep=0pt]
\tikzstyle{hidden}=[circle,draw,minimum size=5pt,inner sep=0pt]
\begin{tikzpicture}
  \node[hidden] (h) [label=above left:$H$]{};
  \node[vertex] (2) at (-1.10,-0.64) [label=above:$X_2$]{};
  \node[vertex] (1) at (0,1.27) [label=above:$X_1$]{};
  \node[vertex] (3) at (1.10,-0.64) [label=above:$X_3$]{};
  \draw[->,-latex,line width=.3mm] (h) to (2);
  \draw[->,-latex,line width=.3mm] (h) to (3);
  \draw[->,-latex,line width=.3mm] (h) to (1);
\end{tikzpicture}
\caption{Tripod tree. A strictly trivalent tree with $n_{l}$ leaves contains $\binom{n_{l}}{3}$ conditional independence relationships which can each be expressed as a tripod tree.} \label{fig:tripod}
\end{figure}
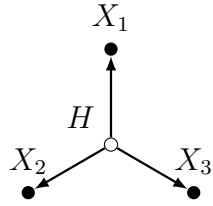

A graph $T$ is a tree if there exists a unique path between any two connected vertices. The trees of interest are those with interior hidden nodes $H \in \mathcal{H}$ and manifest leaf nodes $X \in \mathcal{X}$ (denoted by white and black circles respectively). The observed leaf variables can be thought of as contemporary languages, and the latent interior variables as past versions of languages. Attention is restricted to strictly trivalent trees, i.e. trees where no interior node has more than three adjacent nodes. Any tree from this class with more than three leaves can be expressed as a bifurcating tree, a common model in phylogenetics (e.g.\ \citet{fel78net}). The tripod tree (shown in Figure~\ref{fig:tripod}) describes any such relationship between three observed variables. To see this, simply note that any three connected leaves of a strictly trivalent tree share a unique interior node, and that the conditional independence of these four nodes is minimally described by the tripod tree. Hence, the tripod tree is a fundamental component for analyses involving tree constraints.

The distributions associated with moments of the observed variables of such tree models are known for some settings. The case of univariate binary random variables was expanded in \citet{set00gmc} and has recently attracted considerable interest (for example \citet{all09ipl, zwi12tcg}). These advances have encouraged some authors to proceed to use the known geometry of these spaces to support inference and learning over the space of tree models (see \citet{drt07asm}). These focus on the polynomial constraints that are implicit in these models. \citet{zwi11iic} fully characterized the space for binary trees by extending the understanding beyond algebraic relationships to also include the active semi-algebraic constraints, whilst \citet{all12sdg} broadened the results to a $k$-state $n$-pod tree. Our interest is in the use of the constraints which define these geometries for assessing whether data could have originated from a tree model, in particular whether the spoken languages in the acoustic data set could be modeled as a phylogenetic tree.

\subsection{A Fundamental Constraint on the Covariance of Gaussian Tree Models}
\label{sec:gausscon}

Most of the results in the literature of tree constraints are only applicable in settings with binary random variables. Alternative constraints specific to Gaussian tree models are required in order to perform equivalent tree-amenability diagnostics for Gaussian functional data such as the Romance data set. Here we describe a fundamental univariate Gaussian tree constraint which has not yet been exploited for the purpose of investigating tree-amenability. 

Consider the precision matrix $\Sigma^{-1}$ related to the tripod tree in Figure~\ref{fig:tripod} with one latent and three manifest Gaussian random variables. It is well known in the Gaussian setting that if $X_i$ is independent of $X_j$ conditional on all other observed variables then the corresponding entry $\Sigma^{-1}_{ij} = 0$ (see \citet[Chapter 5]{lau96gm-} for example). So the precision matrix for the univariate Gaussian tripod tree has the form:
$$
\bm{\Sigma}^{-1}=\left(
\begin{array}{ccc|c}
  \sigma^{-1}_{1} & 0 & 0 & \sigma^{-1}_{1H}\\
  0 & \sigma^{-1}_{2} & 0 & \sigma^{-1}_{2H}\\
  0 & 0 & \sigma^{-1}_{3} & \sigma^{-1}_{3H}\\ \hline
  \sigma^{-1}_{1H} & \sigma^{-1}_{2H} & \sigma^{-1}_{3H} & \sigma^{-1}_{H}
\end{array}
\right)
$$
where the final row/column relates to the interior hidden variable. The covariance matrix resulting from taking the inverse of the precision matrix can be expressed algebraically in terms of entries of $\Sigma^{-1}$. Using the resulting entries of the covariance $\Sigma = [\sigma_{ij}]$ it is then straightforward to calculate that a necessary condition for the leaves of this tree to be the margin of the tripod tree given in Figure~\ref{fig:tripod} is that
\begin{equation}
\sigma_{ij}\sigma_{ik}\sigma_{jk} \ge 0 \fourall i < j < k.
\label{eqn:posit}
\end{equation}
We shall refer to this constraint as the positivity constraint. For a strictly trivalent tree with $n_{l}$ observed leaf nodes, there are $\binom{n_{l}}{3}$ tripod trees that must be valid - one for each triple - and thus $\binom{n_{l}}{3}$ such positivity constraints. Whilst there are further constraints imposed on the observed moment space of  trees, the derivations are considerably more involved and not the emphasis of this applied paper. Given that applications of Gaussian tree-constraints are not apparent in the literature, there is still much to explore focusing solely on this fundamental Gaussian tree constraint.

These constraints can be used with the linguistic data set as a diagnostic for assessing tree-amenability. More precisely, for the $r$ component-by-component projections, each can be assessed for tree-amenability. Thus some components may be found to violate the tree constraints whereas others may satisfy them. This may suggest which attributes of the spoken languages have evolved in a tree-like manner and which have not. Note that although we have functional data, the formulation of this positivity constraint does not assume this, and thus the constraint is equally valid for multivariate Gaussian data.


\subsection{Constructing a Suitable Covariance Statistic}
\label{sec:treeapp}

In pursuit of assessing tree-amenability of the $r$ projections of the Romance data using the positivity constraint $\sigma_{ij}\sigma_{ik}\sigma_{jk} \ge 0$, it is clear that a sample covariance of the scores must be constructed. Recall that the relationships of interest in this study are at the language level and thus between-language covariances (each $5 \times 5$) are the appropriate statistics to produce, one for each of the $r$ components. 
One approach to calculating the entries of these matrices is to treat the mean score of each word in a language as an observation and then measure the distance from the overall word mean projection. Then using appropriate weights, a between-language covariance matrix can be estimated as follows. Let $\bar{y}_{d}^i = \frac{1}{m_{\dotr d}}\sum_{l=1}^{n_l}m_{ld}\bar{y}_{ld}^{i}$, $m_{\dotr d}=\sum_{l}^{n_l}m_{ld}$ where recall $m_{ld}$ is the number of samples of word $d$ in language $l$, and $\bar{y}^i_{ld} = \bm{c}_i\bar{x}_{ld}$ the projection of the mean of word $d$ of language $l$ using component $\bm{c}_i$.
Then for component $i$ the between-groups cross-covariance for the projected data has the following form:
\begin{equation}\bm{\Sigma_{Y_{i}}} = [\sigma^{i}_{l,l'}] \text{ where }
\sigma_{l,l'}^{i} =\sum\limits_{d=1}^{n_{d}} \frac{\sqrt{m_{ld}}\sqrt{m_{l'd}}(\bar{y}^i_{ld} - \bar{y}^{i}_{d})(\bar{y}^{i}_{l'd}- \bar{y}^{i}_{d})}{n_{d}-1}
\label{eq:outputbet}
\end{equation}
where recall $n_d$ is the number of unique words. Note that this between-group covariance differs from that used in the CVA - this is of the projected data, with word means treated as observations. This is a valid construction in the sense that (\ref{eq:outputbet}) is an inner product (see \citet{ist87ips} for instance). The sample matrices $\bm{\hat{\Sigma}_{Y_{i}}}$ will be rank deficient if $n_l \geq n_d$. Also, observe that if for at least one word $d$ the number of observations is unequal across languages then the weighted word mean $\bar{y}^{i}_{d}$ differs from the unweighted version. This relaxes a zero-sum condition on the the rows or columns of $\bm{\hat{\Sigma}_{Y_{i}}}$ permitting the covariance matrix to be full rank. In the alternate case of a balanced observational design, full rank can be achieved through an alternative construction (for example adding the unweighted word means back to each language-word mean).

Now component-by-component covariances can be used to indicate adherence to a Gaussian tree model using the tripod tree positivity constraint on all $\binom{n_l}{3}$ selections of languages. Each component captures a different combination of variability. Thus it is not unexpected that some components may show violations of the constraint whereas others may indicate tree-amenability.

\subsection{Simulation}
\label{sec:sim}
Before analyzing the Romance language data, a short simulation is performed. The purpose of the simulation is to demonstrate the effectiveness of the tripod positivity constraint at identifying tree-amenability. Also, by varying the simulation sample size the robustness of the positivity constraint is examined.

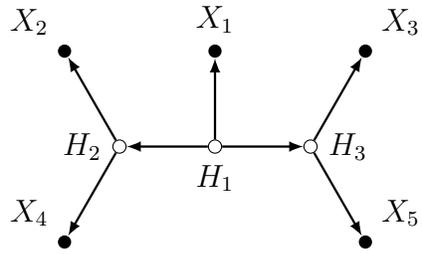
\begin{figure}[ht!]
\centering
\tikzstyle{vertex}=[circle,fill=black,minimum size=5pt,inner sep=0pt]
\tikzstyle{hidden}=[circle,draw,minimum size=5pt,inner sep=0pt]
\begin{tikzpicture}
  \node[hidden] (h1) [label=below:$H_{1}$]{};
  \node[hidden] (h2) at (-1.27,0) [label=left:$H_{2}$]{};
  \node[vertex] (x1) at (0,1.27) [label=above:$X_{1}$]{};
  \node[hidden] (h3) at (1.27,0) [label=right:$H_{3}$]{};
  \node[vertex] (x2) at (-2,1.27) [label=above left:$X_{2}$]{};
  \node[vertex] (x4) at (-2,-1.27) [label=above left:$X_{4}$]{};
  \node[vertex] (x3) at (2,1.27) [label=above right:$X_{3}$]{};
  \node[vertex] (x5) at (2,-1.27) [label=above right:$X_{5}$]{};
  \draw[->,-latex,line width=.3mm] (h1) to (h2);
  \draw[->,-latex,line width=.3mm] (h1) to (h3);
  \draw[->,-latex,line width=.3mm] (h1) to (x1);
  \draw[->,-latex,line width=.3mm] (h2) to (x2);
  \draw[->,-latex,line width=.3mm] (h2) to (x4);
  \draw[->,-latex,line width=.3mm] (h3) to (x3);
  \draw[->,-latex,line width=.3mm] (h3) to (x5);
\end{tikzpicture}
\caption{Five leaf tree.} \label{fig:5pod}
\end{figure}

In order to mimic the linguistic data set, a tree with five leaves is used to generate data (see Figure~\ref{fig:5pod}). Standard structural equations are used (\citet{bol89sel}) adapted for a Gaussian tree to model the dependencies between nodes. The interior node $H_1$ is the root and this variable is simulated from a Gaussian distribution. Data on nodes adjacent to the root are generated as linear combinations of this root simulation with the addition of Gaussian noise. In a $p$-dimensional scenario, the root node is simulated as follows:
$$\bm{H}_{1} \sim \mathcal{N}_{p}(\bm{0}_{p}, \bm{\mathcal{V}}_{h_{1}})$$
where $\bm{\mathcal{V}}_{h_{1}}$ is a $p \times p$ symmetric positive matrix, constructed as $\bm{\mathcal{A}} + \bm{\mathcal{A}}^{T} + p\bm{I}_{p}$ with entries of $\bm{\mathcal{A}}$ generated independently from the continuous uniform distribution $U(0,1)$. $\bm{0}_{p}$ is a $p$-vector of zeros, and $\bm{I}_{p}$ is the $p \times p$ identity matrix. Subsequent nodes are simulated from previous ones, for example:
$$\bm{X}_1 = \bm{\lambda}_{h_{1}x_{1}} H_1 + \bm{\epsilon}_{x_{1}}$$
Each entry of $\bm{\lambda}_{h_{1}x_{1}} \sim U(-a,a)$, $a\in \mathbb{R}$. Within a repetition, all entries of $\bm{\lambda}_{h_{1}x_{1}}$ are fixed for all observations, although do differ depending on the node transition being considered. Additionally, $\bm{\epsilon_{x_{1}}} \sim \mathcal{N}_{p} \left(\bm{0}_{p},b^{2}\bm{I}_{p}\right)$, $b\in \mathbb{R}$, and $\bm{\epsilon}_{x_{1}}$ is pairwise independent with all other random variables. Following the directions of the arrows, the remaining variables are simulated equivalently.

A four-dimensional data set $D$ is generated for all eight nodes of the tree though only the observed leaf nodes are of interest and are utilized with the tree constraint. Another data set $D^{*}$ is also generated using a corrupted version of the tree where each transition between nodes provides an opportunity for sign-reversal of entries of the data. Given this corruption can occur on any dimension of any of the data, the simulation is no longer from a Gaussian tree model. Although higher dimensional data could be generated, under standard CVA only a maximum of four ($n_{l} - 1$) eigenvalues would be non-zero (the use of separable CVA relaxes this assumption in our analysis). Thus there would not be much information gained from the increased dimension but the computational resource required would certainly be greater. Four sample sizes are investigated: 50000, 5000, 500, and 50. These are average sample sizes as the simulation is designed such that the number of observations differs across languages allowing the use of (\ref{eq:outputbet}) for the reasons relating to rank noted in Section~\ref{sec:treeapp}. Each simulated data set undergoes the same basis change using CVA to approximate CFA as described in Section~\ref{sec:cva}. For a full analysis, all four dimensions are retained for projecting the data, and thus for each data set four covariance matrices are calculated using the construction given in (\ref{eq:outputbet}). The resulting covariance matrices are then used to check the positivity constraint $\sigma_{ij}\sigma_{ik}\sigma_{jk} \ge 0$. Having repeated the simulation 1000 times for each sample size the results are summarized in Table~\ref{tab:posd}. For the projections using components $\bm{c}_{1}, \ldots, \bm{c}_{4}$ (ordered by explanatory power), the percentage of samples satisfying the positivity constraint is recorded.

\begin{table}[htb]

\caption{Percentage of simulations which satisfy the positivity constraint for the sample size (left headings) and for the four components for each data set $D$ and $D^{*}$ (above headings), for parameter values $a=5$, $b=6$ in the structural equations.}
\begin{center}
\begin{tabular}{c c c c c c c c c}
\hline
\multicolumn{1}{c}{Sample} & \multicolumn{4}{c}{${D}$} & \multicolumn{4}{c}{${D^{*}}$}\\ 
\multicolumn{1}{c}{Size} & $\mathbf{c_{1}}$ & $\mathbf{c_{2}}$ & $\mathbf{c_{3}}$ & $\mathbf{c_{4}}$ & $\mathbf{c_{1}}$ & $\mathbf{c_{2}}$ & $\mathbf{c_{3}}$ & $\mathbf{c_{4}}$ \\ \hline
\multicolumn{1}{c}{{50000}} & 100 & 100 & 97 & 61 & 11 & 9 & 10 & 10 \\
\multicolumn{1}{c}{{5000}} & 100 & 99 & 85 & 41 & 13 & 7 & 9 & 11 \\
\multicolumn{1}{c}{{500}} & 98 & 87 & 58 & 28 & 13 & 8 & 9 & 12 \\
\multicolumn{1}{c}{{50}} & 88 & 61 & 27 & 27 & 12 & 9 & 11 & 11
\end{tabular}
\end{center}
\label{tab:posd}
\end{table}

There are three notable features highlighted by Table~\ref{tab:posd}. First, the data sets $D$ are found to be tree-amenable consistently more than $D^{*}$. Second, the performance of the positivity constraint on $D$ decreases as the sample size decreases, particularly in the third and fourth components, whereas for $D^{*}$ there is little difference. Third, it is the highly explanatory components of $D$ 
that are most effective at correctly satisfying the positivity constraint. Further investigation into this last feature can give a guide as to how much variance a component should account for before the positivity constraint becomes a reliable diagnostic. Using the $D$ simulations, the proportion of times tree-amenable components are correctly identified as such is estimated over small ranges of the explanatory power; the interpolated estimates are plotted as the midpoints of the ranges. Although the results displayed are for when the structural equation parameter values are $a=5$, $b=6$, similar results are found when these parameters are varied. In general, as $a$ increases and $b$ decreases the performance of the constraint $D$ over $D^{*}$ is more notable.

\begin{figure}[htb!]
\centering
\includegraphics[width=0.85\textwidth, height=0.85\textheight,keepaspectratio]{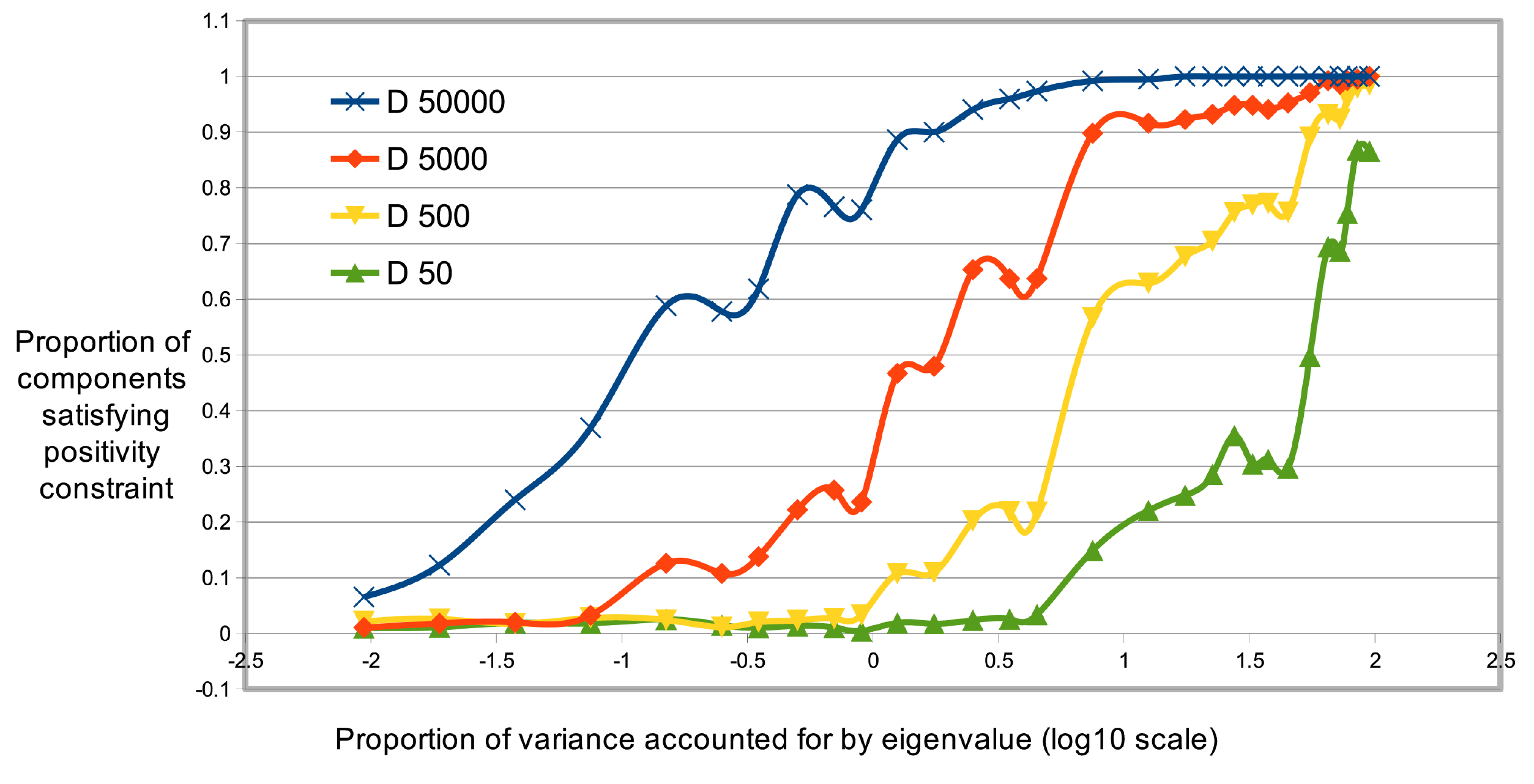}
\caption{For each sample size, interpolated plots are given of the estimated relationship between the explanatory power of components known to be tree-amenable and the chances that the components are correctly identified as being tree-amenable. This broadly suggests that the larger the sample size and/or the higher the explanatory power of a component, the higher the chances of a component being correctly identified as tree-amenable.}
\label{fig:propeigtestcva}
\end{figure}

Figure~\ref{fig:propeigtestcva} displays graphically the property indicated in Table~\ref{tab:posd} - that the lower the explanatory power of a component and the lower the sample size, the less reliable the positivity constraint is as an indicator of tree-amenability. However, the performance of the positivity constraint is still relatively high; even in the lowest sample size, the first component is effective 88\% of the time. Furthermore, the simulations suggest the reliability of the constraint is not symmetric, particularly for low sample sizes; if a component satisfies the positivity constraint then this is appears good evidence of true tree-amenability, but if a component does not satisfy the positivity constraint then underlying tree-amenability should not be ruled out.

\subsection{Further Simulation Based on Characteristics of Observed Data}
\label{sec:furthersim}
In order to better assess the fundamental Gaussian tree constraint when applied to the data, a further simulation is performed using characteristics of the acoustic data set (e.g.\ sample size, eigendecomposition, sample variance). This is achieved by generating data from cross-covariance matrices $\bm{\Sigma_{Y}}$ (as in (\ref{eq:outputbet})) for which tree-amenability status is known. Then by reversing the eigenbasis projection as performed for the Romance data set, new observations are obtained. These new samples can then be assessed using the fundamental Gaussian tree constraint, and it can be assessed whether the source $\bm{\Sigma_{Y}}$ being tree-amenable or not has an affect on the new sample being deemed tree-amenable.

%

Three scenarios are considered: (A) cross-covariance $\bm{\Sigma_{Y}}$ which is tree-amenable; (B) $\bm{\Sigma_{Y}}$ which is not tree-amenable; (C) $\bm{\Sigma_{Y}}$ is tree-amenable 50\% of the time independently for each component and replication. For (A) and (B) the $\bm{\Sigma_{Y_{i}}}$ calculated from the Romance data set are used (for the analysis presented in Section~\ref{sec:apply}). Each of these 15 $\bm{\Sigma_{Y_{i}}}$ are thus known to be tree-amenable or not for the fundamental Gaussian tree constraint, and furthermore, the number of $\binom{5}{3} = 10$ constraints satisfied is known and can be denoted $T(\bm{\Sigma_{Y_{i}}})$. Recall, $\bm{\Sigma_{Y_{i}}}$ is only deemed tree-amenable if $T(\bm{\Sigma_{Y_{i}}}) = 10$. That is, it satisfies the 10 positivity constraints comprising products of covariances relating to the 10 possible choices of three languages. Note that it is not possible for $T(\bm{\Sigma_{Y_{i}}})$ to be exactly 1, 2, 8 or 9 due to the appearance of each off-diagonal entry of $\bm{\Sigma_{Y_{i}}}$ in three of the ten combinations. For the first 15 $\bm{\Sigma_{Y_{i}}}$ from the actual data set, the percentage of tree-amenable samples out 1000 generated is recorded against $T(\bm{\Sigma_{Y_{i}}})$. The four $\bm{\Sigma_{Y_{i}}}$ which are tree-amenable provide higher proportions of tree-amenable samples than the remaining covariance matrices. Considering the full results (see Table~\ref{tab:sim} in the Appendix), a positive correlation is seen between $T(\bm{\Sigma_{Y_{i}}})$ and the percentage of samples simulated from $\bm{\Sigma_{Y_{i}}}$ which are then found to be tree-amenable. Considering scenario (C), a mix of the best and worst performing of the actual $\bm{\Sigma_{Y_{i}}}$ are used, having respectively 35\% and 7\% sample tree-amenability. The simulation results in an overall 18\% tree amenability across all sampled components, suggesting that proportionately a similar level of tree-amenability is retained even when tree and non-tree components are combined.

This second simulation approach closely resembles the linguistic data set both in structure and parameter values. Thus these simulations provide comfort that even for the relatively small sample size the analysis is able to produce reasonable results. Scenarios (A) and (B) provide evidence that components which truly satisfy a low number of constraints are unlikely to give a false positive in terms of tree-amenability. Usefully they also identify that this risk grows as the number of constraints satisfied increases, and this can thus be considered when assessing the results. Scenario (C) demonstrates that tree-amenable components can be recovered even when combined with non-tree components, and furthermore, that these appear to be retained in the output almost proportionally to the input. The sampling in all these simulation scenarios may also provide a proxy for distributional results of tree-amenability, where the higher the proportion of tree-amenability in a sample the more robust the conclusion. This hypothesis is an open problem which we look to address in a subsequent paper through derivation of explicit moments of relevant parameters.


\section{ASSESSING TREE-AMENABILITY OF ROMANCE LANGUAGES}
\label{sec:apply}

As illustrated in Figure~\ref{fig:2dproj}, an effective projection of the Romance data can be performed in even two dimensions. However, to account for a higher proportion of between- to within-language variability further components can be included for the projection. 
A threshold of cumulative between-language variability is set at 97.5\% which for these data equates to including all components which account for at least 0.05\% explanatory power. Application of this threshold provides a dimension reduction from 8100 to 15. Each one of these 15 components $\bm{c}_{1}, \ldots, \bm{c}_{15}$ accounts for some mode of variability between languages. Although the earlier components have high explanatory power, the latter components may isolate directions of variability which are of more interest from a linguistic perspective.

Applying the positivity constraint to each of the 15 component covariance matrices results in four of the components ($\bm{c}_{1}, \bm{c}_{2}, \bm{c}_{4}, \bm{c}_{6}$) adhering to the positivity constraint. Recall that the simulations in Section~\ref{sec:sim} hint that for less explanatory components the constraint appears more likely to identify true tree components than false positives. This suggests that the identified tree-amenable components are more likely to truly be so, whereas the rejected components (e.g. $\bm{c}_{3}, \bm{c}_{5}$) may or may not be truly tree-amenable.

\begin{figure}[htbp]
\centering
\includegraphics[width=1\textwidth, height=1\textheight,keepaspectratio]{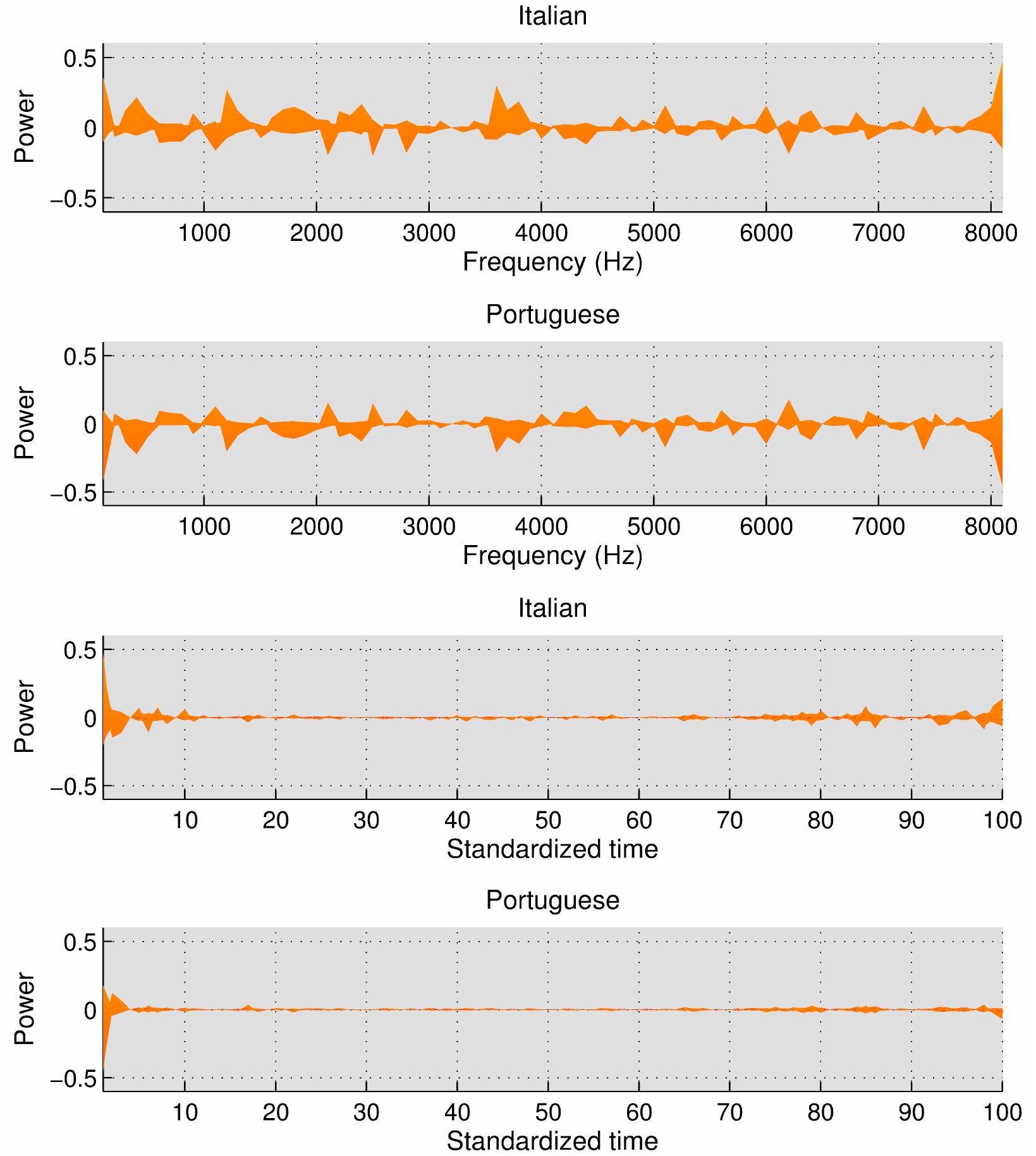}
\caption{Cross-sections from the frequency and pseudo-time perspectives of the interpolated second component Hadamand matrices. For the frequency plots, the points with power of opposite sign (for instance points denoted 3500-4500Hz) indicate the ranges which are separating Italian and Portuguese in the second co-ordinate. The time points with higher opposite powers are between time units 1-20 and 70-100. Thus it is the non-central time points which are useful for distinguishing Italian and Portuguese in the second co-ordinate.}
\label{fig:ftcomp2}
\end{figure}

To develop an insight into which aspects of the languages are being identified by the separable-CVA, the Hadamard (entrywise) product of each component with each concatenated mean language spectrogram is calculated. Restoring the $81 \times 100$ dimensions, the resulting matrices indicate the contribution of each frequency-time point to the overall co-ordinates produced by the component projections. This is useful for highlighting particular ranges of standardized time and frequency which distinguish between languages. 
For illustration, the time and frequency perspectives are plotted for the second component for both Italian and Portuguese in Figure~\ref{fig:ftcomp2}. These two languages are selected as they are clearly separated in both projected co-ordinates (Figure~\ref{fig:2dproj}), although the plots for the remaining languages are similar. Contrasting the plots in Figure~\ref{fig:ftcomp2}, immediately it can be seen that there are many frequency points which contribute to the overall projection whereas in the time dimension most points do not appear to be integral to distinguishing between languages in the second component. Considering the frequency perspectives, there is some symmetry in power between Italian and Portuguese. This is exhibited well in frequency ranges 300-800Hz, 1000-1500Hz, 3500-4500Hz, and 6000-6500Hz which show reflections along the line of zero power. Each of these frequency ranges can be phonetically interpreted. The 300-800Hz range likely relates to the first formant F1 being different, due to vowel differences. The 1000-1500Hz range likely relates to nasality (Portuguese is more ``nasal'' than Italian, and therefore has less energy in this portion of the spectrum than Italian). The 3500-4500Hz range is in the region of the third formant and could correspond to differences in lip rounding between speakers of the languages.  The variation at the highest frequencies, around 6000 Hz, are likely to be due to idiosyncratic differences in speakers (since humans cannot readily control speech frequencies in that range) or in the recordings (equipment or recording location). Examining a smaller range of the data that excludes these high frequencies does not affect the main results of the analyses (data not shown).

Considering the time perspective, it is clear that the interesting time ranges are approximately 1-20 and 70-100. This suggests that the differences in the earliest and latest portions may be particularly effective at separating the languages. These results appear robust to trimming of the data set as well as to standardization of the covariance (data not shown), which suggests that the analysis is identifying more than just a feature of the data registration.

These projections are particularly effective at indicating graphically the dominant features of components which are often obscured when displayed numerically. Of course more detailed analyses are required to determine whether these are general features of the languages or simply of the particular data set studied. However, in either case, this type of exploratory data analysis is clearly a helpful one to perform before any more detailed analysis has taken place, especially if based on the firm assumption that the data originates from a tree.

This application demonstrates a method for isolating and identifying distinguishing aspects of variability in acoustic functional data which may be of evolutionary interest. It shows that it is possible to identify prominent features which render particular components effective for distinguishing the language groups. However, it also highlights the challenge of precise physical interpretation of particular components, a task which appears notably more complex due to having both a time and a frequency dimension. It would be of interest to express these differences back in the sound domain, although given the difficulties in inverting spectrograms to sound, this is not a trivial task. However, it is the subject of ongoing work, including experiments with other parametric acoustic representations that are more easily inverted.

\section{DISCUSSION}
\label{sec:disc}

This paper presents a method for assessing tree-amenability of Gaussian functional data via Gaussian tree constraints. Through simulation the capability of the positivity constraint and its level of robustness to sample size have been demonstrated. Application of this fundamental tree constraint has indicated that the implementation is practically feasible. Moreover, it has shown that meaningful (albeit high-level) interpretations can be made making this particularly useful as an exploratory data tool.

The application to a linguistic data set comprising recordings from Romance language speakers provides several interesting preliminary results. First, it suggests that at least four components of the new basis are tree-amenable, and hence, the linguistic features these particular components represent may have a tree structure. Second, examining projections of the two tree-amenable components with highest explanatory power indicates that the second component is sufficient to distinguish languages, and including the first dimension separates dialects. Finally, closer study of the first two tree components indicates that broadly it is the beginning and end of utterances which are identified as effective language discriminants. Whilst these findings are somewhat speculative and the scope of the analysis is limited to a small set of words, it does provide a starting point for describing interesting tree and non-tree like features of grouped data which may otherwise be obscured. Furthermore, if applied on a larger scale to languages lacking established historical pathways, these techniques may offer fresh indications of the plausibility of different hypotheses concerning their historical development.

In the process of preparing data for tree constraint diagnostics, the concept of decomposable covariance structure has been combined with CVA to produce separable-CVA. Although not central to the aim of the paper, it is worth underlining its usefulness. As with standard CVA it identifies modes of variability which effectively distinguish grouped data, but furthermore, in many circumstances it overcomes the common functional data problem of variable dimension exceeding sample size. This makes separable-CVA an appealing tool to use in practice and one which worked well for separating these languages (Figure~\ref{fig:2dproj}).

Whilst the range of techniques applied in this study can be carried across to other functional data sets, it is important to consider the assumptions which underpin the methods employed. Although no data set will truly obey all of the underlying assumptions, it is important nonetheless to understand the purpose of and the consequences of not satisfying each.

The theory of CVA assumes that the data are Gaussian and that the within-language covariances are equal. Considering the acoustic data, Gaussianity is unlikely to strictly hold. However, it may be sufficiently close to doing so and the first and second order moments may still answer questions about language relationships. If there is information available that the Gaussianity assumption is violated (as may well be evident in larger studies than the one presented here), then applying a copula transformation (see \citet{gen95sep}) will provide us with transformed variables that are marginally Gaussian (a necessary condition for joint Gaussianity). The use of such copula techniques would broaden the scope of these diagnostics and shall be described formally in future work.

In practice the within-language covariances are not likely to be equal (particularly given the sample size of the study). Nevertheless, pooling the covariances will emphasize shared commonality of covariances, and pooling may produce a more stable estimate than otherwise would be possible. Importantly, in circumstances where either assumption is violated CVA still produces a valid basis, the downside being it will not necessarily be as efficient at capturing the variability thus hindering the dimension reduction.

Covariance separability may also be unrealistic as there is often some correlation between the two separated dimensions (e.g.\ frequency and time), but in practice this is not a problem. CVA is an optimality tool and thus deviations from the separability assumption only decrease the efficacy of the basis obtained. That is, if covariance separability holds perfectly CVA determines an ordered basis which linearly maximizes between- to within-language variation as intended. Otherwise the ordered basis is suboptimal, the consequence being more dimensions may be required to capture a suitable proportion of the variance.

Therefore, although the three stated assumptions are not guaranteed to hold, infringement of any of them may impair performance but is unlikely to completely negate such an exploratory analysis. This permits a greater range of data sets for which a similar study could be applicable.

Aside from improving performance by selecting a data set which better meets the assumptions outlined above, the application to the Romance languages could instead be improved through a larger data set with greater breadth of words. Currently, the between-covariances of the projected data are only calculated using the means of the ten unique word observations. Furthermore, some languages have more recordings than others. By increasing the number of recordings for all groups, whilst also including a wider variety words, the outcomes of the analyses are likely to be more robust, with the caveat that inclusion of new words must be carefully considered to ensure they are linguistically suitable across all languages.

In this paper we check only some of the conditions necessary for consistency with a phylogenetic tree. However, it is possible to enhance the power of such diagnostics to include all such checks of possible violations of phylogenetic tree constraints. One method is to utilise the 4-point conditions based on the metric properties noted in \citet{bun71rtm}. If these conditions are satisfied then the tree structure can be uniquely identified. For one particular constraint based on tetrads, the work of \citet{bol93cta} gives a basis for possible test statistics and \citet{drt08mmw} provides some useful distributional results for particular moments. An exploratory approach could use an extension of the binary graphical tree diagnostics described in \citet{shi12gid} to the Gaussian domain and is currently under investigation. Of course to fully and formally evaluate these methods it will be necessary to evaluate the probabilistic properties of our methods. We are currently investigating a Bayesian approach where replicates are generated from Wishart covariance matrices, with properties derived from the projected data. These then provide posterior distributions across a graphical model space (\citet{ata05mcm}). For graphs with high posterior probability, trees with hidden variables could then be induced. As with any formal testing framework, it will be important to consider potential effects of multiple comparisons, and to adjust appropriately for them (see \citet[Chapter 12]{how12smp}).

Each of these methods could enrich the analysis, as the proposed trees could be compared to existing knowledge about the relationships between the languages. Contrasting results could even suggest new directions of research using traditional linguistic methods.


There is much potential for applications in fields beyond linguistics to make use of additional algebraic and semi-algebraic constraints. However, even existing constraints are currently underutilized. For example, despite the relatively simple derivation of the fundamental Gaussian tree constraint, this study marks its first use as a diagnostic for tree-amenability. The inclusion of further tree constraints, both in univariate and multivariate settings could provide further insight into phylogenetic relationships in applications. However, derivation and application of such constraints is non-trivial and is the subject of ongoing work.



\newpage
\renewcommand{\baselinestretch}{1.2} \rm

\section*{APPENDIX: TABLE OF SUMMARISED SIMULATION RESULTS}
\label{app:A}
 \setcounter{table}{0}
  \renewcommand{\thetable}{A.\arabic{table}}
\begin{table}[h!]
\caption{Percentage of components which are tree-amenable when simulated for the different scenarios discussed in Section~\ref{sec:furthersim}.}
\begin{center}
\begin{tabular}{ c c c c c}
\hline
\textbf{Scenario} & \textbf{\begin{tabular}[c]{@{}c@{}}Component\\ ($i$)\end{tabular}} & \textbf{\begin{tabular}[c]{@{}c@{}}Explanatory power\\ of component (\%)\end{tabular}} & \textbf{$T(\Sigma_{Y_{i}})$} & \textbf{\begin{tabular}[c]{@{}c@{}}\% of simulated \\ components\\ tree-amenable\end{tabular}} \\ \hline
(A) & 1  & 94.82  & 10 & 35.1 \\
(A) & 2  & 0.88 & 10 & 22.1 \\
(B) & 3  & 0.43 & 7  & 18.3 \\
(A) & 4  & 0.30 & 10 & 27.6 \\
(B) & 5  & 0.27 & 6  & 16.2 \\
(A) & 6  & 0.17 & 10 & 23.4 \\
(B) & 7  & 0.10 & 4  & 6.9  \\
(B) & 8  & 0.08 & 7  & 10.7 \\
(B) & 9  & 0.08 & 6  & 16.3 \\
(B) & 10 & 0.07 & 5  & 8.6  \\
(B) & 11 & 0.07 & 5  & 10.5 \\
(B) & 12 & 0.06 & 4  & 8.3  \\
(B) & 13 & 0.05 & 4  & 8.4  \\
(B) & 14 & 0.05 & 6  & 12.2 \\
(B) & 15 & 0.05 & 4  & 7.8  \\
(C) & 1 and 7  & 97.25 and 0.10 & 10 and 4 & 17.8
\end{tabular}
\end{center}
\label{tab:sim}
\end{table}
\renewcommand{\baselinestretch}{1.63} \rm

 In general, the higher the value of $T(\Sigma_{Y_{i}})$, the higher the percentage of tree-amenable components in the sample. Scenario (C) is a  sample from a mixture of components 1 and 7, the resulting percentage of tree-amenability being half that of component 1 individually.

\bibliographystyle{apalike}
\bibliography{applied}


\end{document}